\documentclass[twocolumn,aps,a4paper,showpacs,amsfonts,amsmath,amssymb]{revtex4}

\usepackage{graphicx}

\begin{document}

\title{Scaling properties and universality of first-passage time probabilities in financial markets}
\author{Josep Perell\'o, Mario Guti\'errez-Roig and Jaume Masoliver}
\affiliation{Departament de F\'{\i}sica Fonamental, Universitat de
Barcelona.\\ Diagonal 647, E-08028 Barcelona, Spain}
\begin{abstract}
Financial markets provide an ideal frame for the study of crossing or first-passage time events of non-Gaussian correlated dynamics mainly because large data sets are available. Tick-by-tick data of six futures markets are herein considered resulting in fat tailed first-passage time probabilities. The scaling of the return with the standard deviation collapses the probabilities of all markets examined, and also for different time horizons, into single curves, suggesting that first-passage statistics is market independent (at least for high-frequency data). On the other hand, a very closely related quantity, the survival probability, shows, away from the center and tails of the distribution, a hyperbolic $t^{-1/2}$ decay typical of a Markovian dynamics albeit the existence of memory in markets. Modifications of the Weibull and Student distributions are good candidates for the phenomenological description of first-passage time properties under certain regimes. The scaling strategies shown may be useful for risk control and algorithmic trading.
\end{abstract}

\maketitle

\section{Introduction}

The first-passage time (FPT) of a given stochastic process is a random variable representing the instant of time when the process first attains some preassigned ``critical'' value. The study of problems related to first-passage time has a long tradition in many branches of science and engineering~\cite{redner,bunde}. In recent years, empirical as well as theoretical analysis on this kind of problems have gained a broader interest. Thus new multidisciplinary contexts like ionic transport in a bacterial porin~\cite{faraudo}, protein folding in a crowded cell environment~\cite{lois}, epidemics spreading in human diseases or computer viruses and human mobility~\cite{barabasi} require a substantial analysis of FPT events. In many cases the customary assumption of an underlying uncorrelated Gaussian diffusion dynamics is no longer applicable~\cite{condamin,klafter}. This is certainly the case of financial markets because market fluctuations are away from being Gaussian owing to fat-tails and the appearance of clustering structures with time~\cite{mantegna,lo}.

Unfortunately, the observation of first-passage time phenomena requires large databases in order to get reliable statistics of the most extreme and, hence, rare events. However, high-frequency financial time series can be truly large usually containing millions of observations. This allows a rather  complete estimation of FPT probability distributions and their tail decay (the latter ruling extreme events) going thus beyond common approaches based only on the evaluation of first moments, {\it i.e.}, the mean first-passage time and the mean exit time \cite{mp}. 

An additional focus of interest is linked to the financial trading industry. The most extreme and less probable situations become crucial to calibrate default probabilities and hence obtaining alternative or improved credit risk estimations~\cite{rutowski}. Less dramatic situations involving the pricing of American and other exotic options also need to measure statistics related to FPT events~\cite{alili,avram}. Moreover, intraday algorithmic trading strategies~\cite{chriss} of buying or selling a given asset could learn from FPT statistics. Computer based trading is thought to be responsible for as much as 73\% of trading volume in the US in 2009~\cite{hendershot}. Yet empirical FPT probabilities are quite unexplored in financial markets~\cite{masoliverfpt,sazuka} although there are some works studying other extreme-time statistics \cite{mp,havlin1,wang,lillo,bonano,valenti,bunde1,bogachev,kantz,ren}. 

The two main goals of this paper are: (i) the estimation of FPT probabilities for a wide class of markets and (ii) their subsequent interpretation and possible classification as universal properties of the market as a whole. We will thus look at financial databases in a high-frequency resolution and thoroughly study the FPT statistics by taking transaction to transaction data (not mid-price) of six futures contracts. These are: DeutscheMark-US Dollar foreign currency (USDM, 04/01/1993--09/12/1997), Standard \& Poor's 500 index (SP500, 04/01/1993--09/12/1997), Deutscher Aktiken index (DAX, 2007/02/13--2007/06/13), Dow Jones Industrial Average Index (DJI, 2006/03/01--2007/08/27), Euro-US Dollar foreign currency (EURUSD, 2007/08/01--2007/08/27), the Spanish index (IBEX, 2007/01/02--2009/12/30). Each database contains millions of transactions. For example, the IBEX data contains 4,613,250 non-simultaneous transactions. We are taking the nearest expiry future contract and look at intraday price statistics. We are excluding first and last 30 minutes of each trading day in order to avoid opening and closing anomalous effects. We are also excluding big jumps due to the roll on to the next nearest expiry future contract.

\section{Analysis}

We will analyze the return defined as the logarithmic price change at time $t$: 
\begin{equation}
X(t)=\ln[S(t)/S_0],
\end{equation}
where $S_0=S(0)$ is the price at the starting time $t=0$. 

\subsection{Scaling across different markets and time horizons}

As stated above, our main objective is estimating the FPT probability distribution, $W(x,t)$, which is defined as the probability that the first-passage time to a target level $x$ is less than $t$. In other words, $W(x,t)$ is the probability that the return first crossed level $x$ before instant $t$:
\begin{equation}
\label{Wxt}
W(x,t)=
\begin{cases}
{\rm Prob}[X(t')>x, 0<t'<t] & \text{if } x>0 \\
{\rm Prob}[X(t')<x, 0<t'<t] & \text{if } x<0.\\
\end{cases}
\end{equation} 

\begin{figure}
\includegraphics[width=8.75cm]{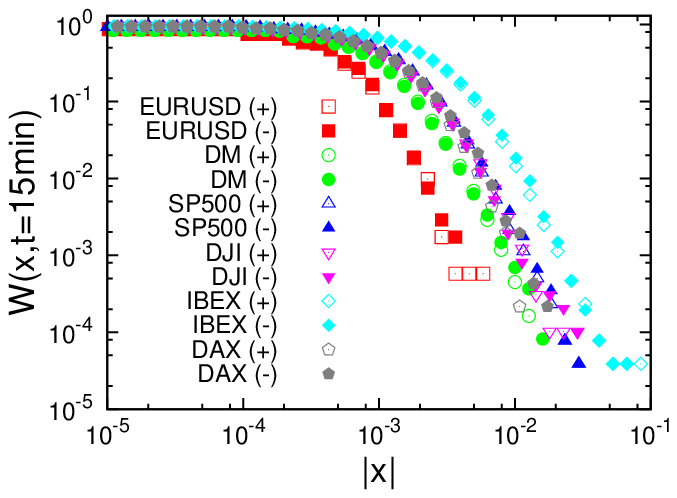}
\includegraphics[width=8.75cm]{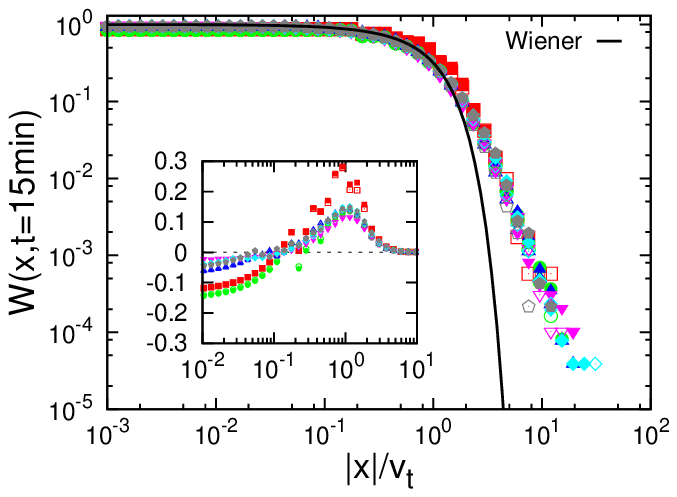}
\caption{(Color online) Log-log representation of FPT probabilities, $W(x,t)$, for different assets as a function of return $x$ (top) and as a function of the scaled return $x/v_{t}$ where $v_t$ is defined in Eq. (\ref{volatility}) (down) when $t=15$ min. We label positive target values, $x>0$, with (+) and negative ones, $x<0$, with (--). Solid line corresponds to the Wiener case~(\ref{wwiener}). The inset in the down figure shows the difference $W(x,t)-W_G(x,t)$ with the FPT probability of the Wiener process $W_G(x,t)$.}
\label{fig1}
\end{figure}

Let us first focus on the behavior of the FPT probability across the different markets selected. Figure~\ref{fig1} shows $W(x,t)$ for different markets at a time horizon $t=15$ min. Very similar results are observed for shorter and longer times. The discounting in futures contracts with respect the underlying spot price or the long-term trending behaviour of futures could make look different the positive ($x>0$) and the negative ($x<0$) wings of the FPT probabilities but these effects are not strong enough observed in our results since we mostly focus on intraday time horizons. Great differences between positive target values and negative target values become noticeable only for much longer time horizons~\footnote{See also Ref.~\cite{masoliverfpt} for observing a very big difference between the two tails when $t=22$ days.}. 

If we scale the target returns $x$ of each market with its own standard deviation:
\begin{equation}
v_t=\sqrt{\langle X(t)^2\rangle - \langle X(t) \rangle^2},
\label{volatility}
\end{equation}
then Fig.~\ref{fig1} shows a fairly neat collapse into a single FPT probability curve after plotting $W(x,t)$ in terms of the scaled returns $x/v_t$. We observe that this is so despite the highly non-Gaussian character of data. We remind at this point that the Wiener model
\begin{equation}
dX_{G}(t)=\sigma dW(t),
\end{equation}
where $dW(t)$ is a Gaussian stochastic process with zero-mean and variance given by $dt$, has the following FPT probability~\cite{redner}:
\begin{equation}
W_{G}(x,t)=\mbox{erfc}\left(x/\sqrt{2\sigma^2 t}\right).
\label{wwiener}
\end{equation}

\begin{figure*}[t]
\includegraphics[width=17.5cm]{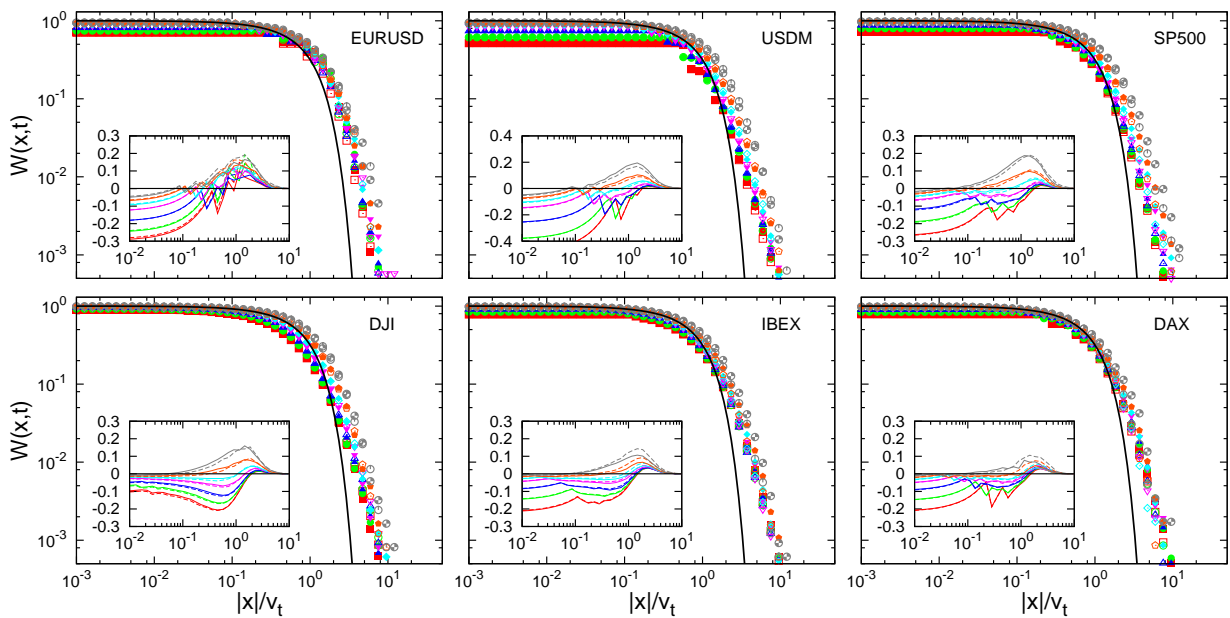}
\caption{(Color online) Log-log representation of the FPT probabilities as a function of the scaled return and with different times. Solid line represents the Wiener case given by Eq.~(\ref{wwiener}). Insets show the difference $W(x,t)-W_G(x,t)$.}
\label{fig2}
\end{figure*}

Notice that scaling enhances the distinction between the empirical results and the corresponding Wiener model with unit variance (see the solid line in the right frame of Fig.~\ref{fig1}). However, collapse still holds for target levels as high as 10 times the size of the (empirical) $15$-minutes standard deviation. The same behavior appears for shorter and longer time horizons (from 1 to 120 minutes) as shown in Fig.~\ref{fig2}. 

The inset in Fig.~\ref{fig1} shows the difference $W(x,t)-W_G(x,t)$ for each market as a function of the scaled return $x/v_t$. As we can see the Wiener model systematically underestimates the empirical FPT probability for large target values ($x \gtrsim v_t$) but somewhat overestimates $W(x,t)$ for smaller levels ($x\lesssim 0.1 v_t$). This remarkable result had been theoretically  predicted by two of us using more sophisticated models such as the CIR-Heston stochastic volatility model~\cite{masoliverfpt}. We believe that this fact may have non-trivial consequences in risk management~\cite{lo,rutowski}.

\begin{figure*}[t]
\includegraphics[width=17.5cm]{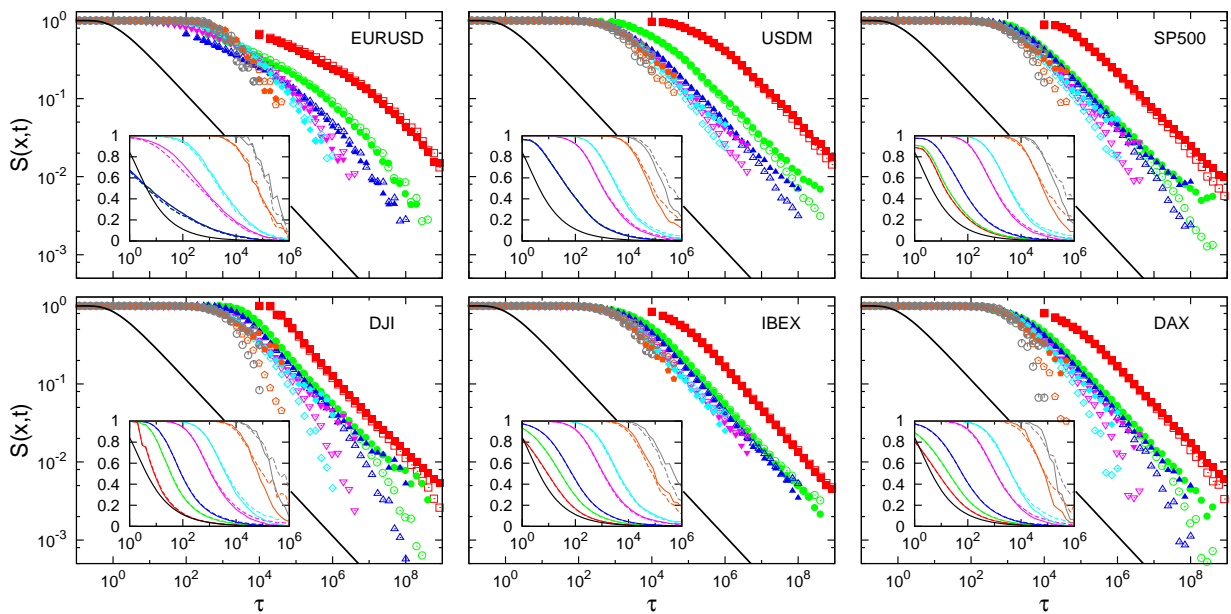}
\caption{(Color online) Log-log representation of the survival probabilities as a function of an scaled time $\tau=(v_{0}/x)^2t$ and for several target values $x$ measured in units of the $30$-minutes standard deviation $v_0$. The insets show the same results with time measured in seconds and in a semi-log scale. Solid line represents the Wiener case when $x=v_0$.}
\label{fig3}
\end{figure*}

Our next step is to select a particular market and observe how the FPT probability changes with time. Figure~\ref{fig2} shows market data for different values of $t$, from $1$ to $120$ minutes. For large time horizons (up to 2 hours), the insets of Fig.~\ref{fig2} show that when $x<v_t$ the underestimation of the Wiener model tends to disappear. In contrast, when $x>v_t$ the overestimation persists and becomes more pronounced as time increases. It is also worth noticing that low resolution in price changes (as is the case of USDM) is also undervaluing $W(x,t)$ when $x/v_t\ll 1$. 

Let us mention that the relative collapse shown in Fig.~\ref{fig2} indicates that, for high frequency data, the tail of the crossing probability $W(x,t)$ is, when properly scaled, independent of the time horizon. Moreover, the fact that this collapse is shared by all markets herein considered strongly suggests that $W(x,t)$ is market independent as shown in Fig.~\ref{fig1} for the time horizon of $15$ min. 

\subsection{Scaling across different target levels}

Another interesting and alternative way of observing extreme-time events is to look for the the survival probability (SP), that is, for the probability $S(x,t)$ {\it of not having reached} level $x$ before time $t$:
\begin{equation}
\label{Sxt}
S(x,t)=
\begin{cases}
{\rm Prob}[X(t')<x, \ 0<t'<t] & \text{if } x>0 \\ 
{\rm Prob}[X(t')>x, \ 0<t'<t] & \text{if } x<0.
\end{cases}
\end{equation}

Note that the SP is related to the FPT probability by the simple relation $S(x,t)=1-W(x,t).$ Figure~\ref{fig3} shows how the SP decays with time for several target levels measured in $30$-minute standard deviation units $v_0$. As shown in Fig.~\ref{fig3}, a collapse is still possible as long as we use the dimensionless time 
\begin{equation}
\tau=(v_{0}/x)^2t.
\label{scaled_time}
\end{equation}
The collapse holds for not very small levels. The smallest level, $x=0.01v_0$, is affected by the presence of finite tick size and by the price autocorrelation~\cite{mantegna,lo}. Additionally, the lack of data affects the largest levels or the data markets with a shorter period of time. See the case of EURUSD data which is due to holding a too short time period (26 days).

The SP decays as $t^{-1/2}$ except for the EURUSD case. This hyperbolic behavior with time also holds for the Wiener model (see Eq.~(\ref{wwiener})) although the latter systematically and drastically underestimates large time horizons. In fact, the hyperbolic decay behavior is the one predicted by the Sparre-Andersen theorem which holds for any symmetric Markov process including L\'evy flights~\cite{redner,sparre,sparre1,chechkin}. The theorem poses now the question about how important are correlations in our data. Transaction to transaction has been then shuffled in two different ways. A first choice randomizes the ordering of return changes while keeping the inter-transaction times as they appear in original data. A second choice consists in randomizing the ordering of inter-transaction times while keeping the original structure consecutive price return changes.

\begin{figure*}[t]
\includegraphics[width=17.5cm]{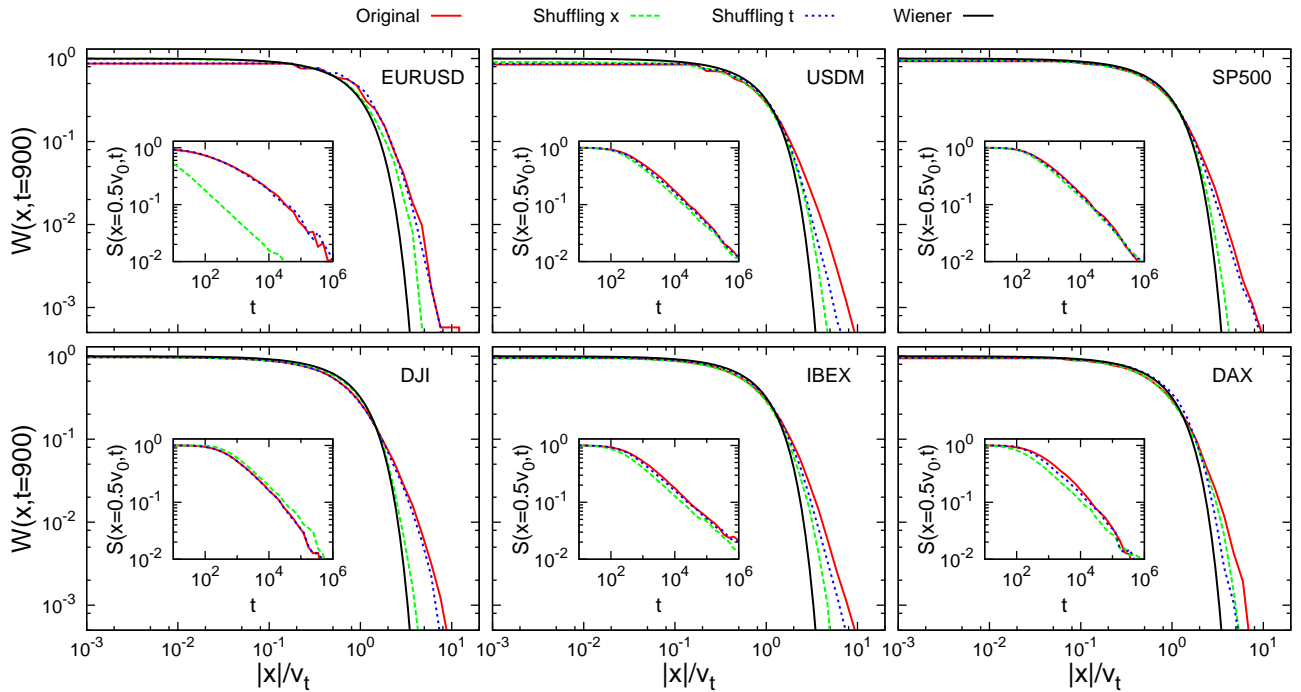}
\caption{(Color online) Log-log representation of the FPT probabilities after shuffling price increments (green dashed line) and interevent times (blue dotted line) in comparison with original curve (red solid line) and Wiener case given by Eq.~(\ref{wwiener}) (black solid line). For the sake of clarity we take only the $900$ seconds time window case when varying $x$ target and $0.5v_{0}$ case when varying $t$ in the inset.}
\label{fig4}
\end{figure*}

Figure~\ref{fig4} shows the resulting FPT distributions after applying the proposed shuffles in returns and in the inter-transaction times. It shows that these two different manipulations lead to two different effects. The randomization in price returns ordering clearly affects the FPT statistics when varying threshold distance. In this case, the tail of the $W\left(x,t\right)$ distribution drastically diminishes to almost the exponential form while the $S\left(x,t\right)$ curve keeps the same hyperbolic decay when varying time horizon. On the other hand, shuffling the inter-transaction times have little effect to most of the markets and keep the extreme statistics curves almost identical. These results are similar in the other time horizons and other target returns studied above. The conclusion that price returns ordering are more relevant than inter-transaction time ordering in the extreme times analysis is consistent with a previous work of two us in the context of the Mean Exit Time~\cite{lillo}.

\subsection{Measuring the dispersion of each collapse}

\begin{table}
\begin{tabular}{lccccc} 
\hline
\hline
&\multicolumn{2}{c}{$\Theta_{x}$ in Fig.~\ref{fig2}}&&\multicolumn{2}{c}{$\Theta_{t}$ in Fig.~\ref{fig3}}\\
&(+)&(--)&&(+)&(--)\\
\cline{2-3} \cline{5-6}
EURUSD & $0.0012$ & $0.0012$ && $0.1150$ & $0.0858$\\
USDM & $0.0023$ & $0.0020$ && $0.1299$ & $0.1772$\\
SP500 & $0.0019$ & $0.0020$ && $0.1181$ & $0.1778$\\
DJI & $0.0018$ & $0.0018$ && $0.1042$ & $0.1829$\\
IBEX & $0.0011$ & $0.0014$ && $0.1831$ & $0.1396$\\
DAX & $0.0007$ & $0.0010$ && $0.0871$ & $0.1542$\\
\hline
\hline
\end{tabular}
\caption{\label{tab1}Dispersion of collapsed FPT and SP curves shown in Figs.~\ref{fig2} and~\ref{fig3} using the form described by Eq.~(\ref{dispersion}).}
\label{table1}
\end{table}

We will now give a quantitative measure for the quality of each collapse by means of a dispersion relation among the different curves. Taking FPT data of a given market at different time horizons (see Fig.~\ref{fig2}) we define the following dispersion measure:
\begin{equation}
\Theta_{x} = \frac{1}{L} \sum_{i=1}^{N} \Delta x_i\sqrt{\frac{1}{M}\sum_{j=1}^{M}\left[W\left(x_{i},t_{j}\right)-\mu_{x_i}\right]^{2}},
\label{dispersion}
\end{equation}
where $M=7$ is the number of different time horizons considered, $\mu_{x_i}$ is the quantity
\begin{equation}
\mu_{x_i}=\frac{1}{M}\sum_{j=1}^M W\left(x_{i},t_{j}\right),
\end{equation}
$\Delta x_i$ is the bin width, $N$ is number of bins and $L=N\Delta x_i$ is the largest target value $x$ considered.

We can also perform a similar evaluation using the SP data shown in Fig.~\ref{fig3} with a dispersion $\Theta_{t}$ defined accordingly:
\begin{equation}
\Theta_{t} = \frac{1}{T} \sum_{j=1}^{M} \Delta t_j\sqrt{\frac{1}{N}\sum_{i=1}^{N}\left[W\left(x_{i},t_{j}\right)-\mu_{t_j}\right]^{2}},
\label{dispersion}
\end{equation}
where $N=7$ is the number of different target values considered, $\mu_{t_j}$ is the quantity
\begin{equation}
\mu_{t_j}=\frac{1}{N}\sum_{i=1}^M W\left(x_{i},t_{j}\right),
\end{equation}
$\Delta t_j$ is the bin width, $M$ is number of bins and $T=M\Delta t_j$ is the largest time horizon $t$ considered.
The two measures of dispersion are summarized in Table~\ref{tab1}. 

We can also evaluate the dispersion for the collapse of the several markets shown in Fig.~\ref{fig1} (right). In this case, and using an equivalent dispersion measure as that of Eq. (\ref{dispersion}), we have $\Theta_{mkt}^+=7.12\times10^{-4}\ (x>0)$ and $\Theta_{mkt}^-=7.99\times10^{-4}\ (x<0).$ We see that
\begin{equation}
\Theta_{mkt}\ll \Theta_{x}\ll \Theta_{t},
\end{equation}
which clearly proves that the best collapse is between different markets and the worst is between time horizons of the SP, as otherwise expected from the visual perception of Figs.~\ref{fig1}--\ref{fig3}.

\subsection{Phenomenological expressions}

\begin{figure}
\includegraphics[width=8.75cm]{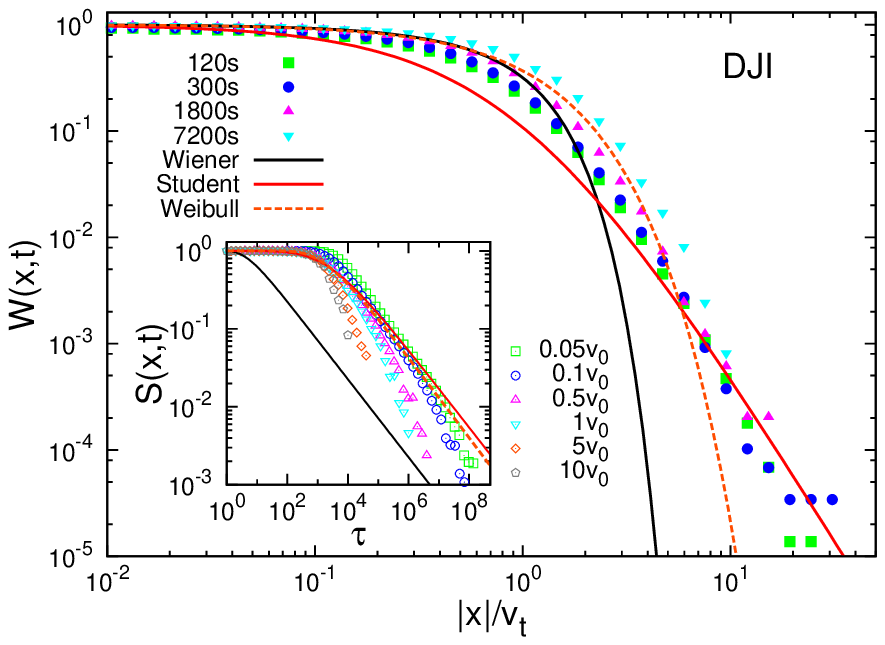}
\includegraphics[width=8.75cm]{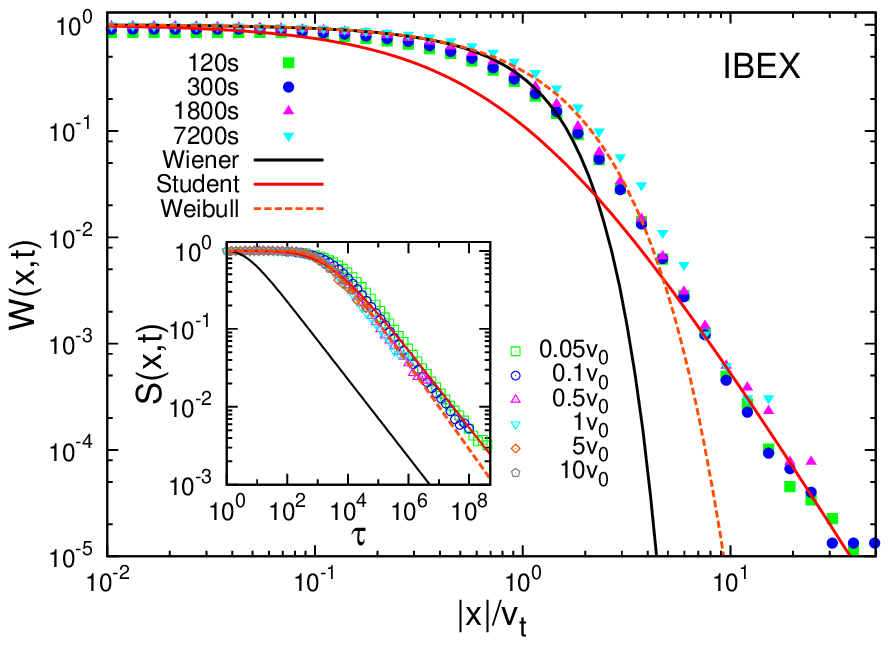}
\caption{(Color online) FTP probability and SP fits of DJI and IBEX using Weibull~(\ref{weibull}) and Student~(\ref{student}) functions in terms of scaled return and scaled time (insets).}
\label{fig5}
\end{figure}

We finish by presenting two phenomenological expressions for the FPT probability aimed to reproduce empirical observations. A first mathematical expression is provided by the Weibull distribution which has been typically a good candidate for adjusting extreme events in a large variety of data sources~\cite{sazuka,klein}. We propose the following modified distribution
\begin{equation}
W_{\rm wei}(x,t)=e^{-\left(x/\sqrt{bt}\right)^\beta},
\label{weibull}
\end{equation}
where the modification to the Weibull's standard form consists in the ad-hoc addition of a square root term which gives the time dependence of the distribution; a term added to reproduce the observed decay in empirical data (see Fig.~\ref{fig3}). In Fig.~\ref{fig5} we check this adjustment together with the following modified (again with a square root term) Student distribution:
\begin{equation}
W_{\rm stu}(x,t)=\left(1+x/\sqrt{at}\right)^{-\alpha}.
\label{student}
\end{equation}

The IBEX and DJI results for positive returns are shown in Fig.~\ref{fig5}. We then use the same parameters values to plot both the FPT and the SP curves (inset) being fairly consistent. Table~\ref{tab2} shows the fitting parameters for all data sets considered. Note that all Weibull exponents are close to $\beta=1$ (the mean exponent of all markets is $\bar\beta=1.04$) which corresponds to the Laplace distribution. Also all power-law exponents of the Student distribution are rather similar: a little above $\alpha=3$ (the mean exponent is $\bar\alpha=3.32$) which is very close to the decay exponent for the market unrestricted probability (not the FPT probability) proposed by Plerou and Stanley \cite{plerou}.

\begin{table}
\begin{tabular}{lccccc} 
\hline
\hline
&\multicolumn{2}{c}{$W_{\rm wei}(x,t)$}&& \multicolumn{2}{c}{$W_{\rm stu}(x,t)$}\\
&$\beta$&$b\times10^{-4} s^{-1}$ &&$\alpha$&$a\times10^{-4} s^{-1}$\\
\cline{2-3}\cline{5-6}
{\scriptsize EURUSD} & $0.93\pm0.01$ & $2.87\pm0.18$ && $3.49\pm0.03$ & $24.9\pm1.1$\\
USDM & $0.98\pm0.01$ & $1.18\pm0.05$ && $3.28\pm0.01$ & $9.4\pm0.6$\\
SP500 & $1.01\pm0.02$ & $1.31\pm0.05$ && $3.49\pm0.01$ & $12.2\pm0.9$\\
DJI & $1.03\pm0.01$ & $1.10\pm0.05$ && $3.21\pm0.02$ & $8.7\pm0.8$\\
IBEX & $1.10\pm0.01$ & $1.08\pm0.01$ && $3.15\pm0.02$ & $8.4\pm0.6$\\
DAX & $1.21\pm0.01$ & $1.05\pm0.02$ && $3.32\pm0.02$ & $9.7\pm0.7$\\
\hline
\hline
\end{tabular}
\caption{\label{tab2}Parameters of the Weibull~(\ref{weibull}) and Student~(\ref{student}) distributions simultaneously fitting FPT probability and SP for $x>0$.}
\end{table}

Finally, Fig.~\ref{fig5} clearly shows, as otherwise expected, that the Student distribution better describes the tails of the empirical curves than it does the Weibull distribution while the situation is reversed for smaller returns, where Weibull performs better than Student, the crossover being approximately at $|x|\sim 5 v_t$. Therefore, models proposed in future to explain FPT statistics should follow this very schematic for small (Weibull) and large (Student) target levels. As to the unrestricted return distribution case, it has not been possible to fit a single distribution for the whole curve.

\section{Conclusions}

To conclude, even though it is a rather unexplored context, financial markets provide an ideal frame for the study of first crossing events of non-Gaussian correlated dynamics essentially because large data sets are available. We have worked on transaction-to-transaction data of six different futures markets and have studied the form of the empirical FPT probabilities. We have obtained different scaling strategies that allow us to collapse these otherwise scattered probabilities into a single curve. 

As a function of the target return $x$, the FPT probability $W(x,t)$ depends not only on the time horizon $t$ but also on the market selected. This triple dependence yields different scaling strategies. The first and most effective scaling corresponds to the FPT collapse, at a given time window and varying threshold distance, across the six different markets surveyed (Fig. \ref{fig1}). This is done by normalizing target returns with their own volatility, calculated as mentioned before. It is then possible to see that, as likewise expected, Gaussian diffusion models underestimate the probabilities of large returns. However, and quite unexpectedly, they overestimate empirical FPT probabilities for small target returns. We believe that this should have substantial practical implications in risk management and control.

A second scaling attempt is made by selecting a single market and looking at the dependence of $W(x,t)$ on the target return for different time horizons. A fairly good collapse of probabilities is observed when returns are again scaled by the standard deviation corresponding to each time horizon. This second scaling allow us to observe how the previously mentioned overvaluation and undervaluation changes with time (see inset of Fig. \ref{fig2}) but has essentially the same behavior in the tail.

The third and last scaling strategy is addressed to the survival probability. In this case we set the target return $x$ at some value and observe the time evolution of the SP. Collapsing curves are then obtained by a time scaling provided by diffusion theory (cf. Eq. (\ref{scaled_time})) even though diffusion behavior is not sustained by empirical data. Looking at Fig. \ref{fig3} we see that the SP collapse works properly, showing the diffusive hyperbolic decay $t^{-1/2}$, only if target returns are neither very small nor very large compared to their standard deviation. A little thought shows that this is the expected behavior since, as is well known, markets are approximately Gaussian away from the tails~\cite{lo,mantegna1} and the center of the distribution~\cite{mantegna1,mmporra}.

The Sparre Andersen theorem~\cite{sparre,sparre1} claims that Markovian processes including the L\'evy flights have a SP with the hyperbolic decay $t^{-1/2}$. We have performed the exercise of shuffling our transaction-to-transaction data in two different ways with the aim of breaking memory in our market data. Figure~\ref{fig4} shows that shuffling in price changes ordering have a much more important effect than inter transaction time ordering. The return suffling drastically diminishes the tail decay of the FPT statistics (cf. Fig.~\ref{fig4}). The hyperbolic decay $t^{-1/2}$ of the SP is preserved after the two shuffling methods.

Based on the scaling strategies obtained, we study modified versions of the Weibull and Student distributions for a phenomenological description of the empirical observations. We have seen that for large returns FPT probabilities are better adjusted by a Student distribution, while probabilities corresponding to intermediate and small returns are better described by a Weibull (almost Laplacian) distribution. This should come as no surprise, since the Student distribution possesses fat tails while that of Weibull does not. Moreover, for the markets herein considered, tail exponents of the Student distribution are tightly packed around their mean value $\bar\alpha=3.32$, which is very near to the tail exponent $\alpha=3$ obtained by Plerou and Stanley~\cite{plerou} for the unrestricted probability of several unrelated markets. This seems to indicate a kind of universal behavior of markets not distinguishing among restricted probabilities (i.e., FPT distributions) and unrestricted ones. 

Let us finally remark two additional universality aspects that may be shared by all financial markets. First, note that under the assumption that data is adjusted by a Weibull or a Student distribution, or by a mixture of them, our scaling strategies suggest the possibility of obtaining the whole first-passage time probability by only estimating the volatility with no need of any other additional data. Whether or not this can be extended to any market should be extensively checked on empirical data and this is beyond the bounds of the present work. Second, and following this somewhat speculative reasoning, we also note that the neat collapse of the FPT probabilities for different markets (see Fig. \ref{fig1}) is a substantial indication that extreme events statistics are, in high-frequency data, market independent. This guess, which sustains a truly universal property, should also be checked on wider markets and wider periods of time.

\acknowledgments
Financial support from Direcci\'on General de Investigaci\'on under contract FIS2009-09689 is acknowledged.


\begin{thebibliography}{30}
\bibitem{george_book} G.H. Weiss, {\it Aspects and Applications of the Random Walk} (North-Holland, Amsterdam, 1994). 
\bibitem{redner} S. Redner, {\it A Guide to First-Passage Processes} (Cambridge University Press, Cambridge, 2001).
\bibitem{bunde} A. Bunde, J. Kropp, and H. J. Schellnhuber, {\it The Science of Disasters} (Springer, Berlin, 2002).
\bibitem{faraudo} C. Calero, J. Faraudo, and M. Aguilella-Arzo, Phys. Rev. E {\bf 83}, 021908 (2011).
\bibitem{lois} G. Lois, J. Blawzdziewicz, and C.S O'Hern, Phys. Rev. E {\bf 81}, 051907 (2010).
\bibitem{barabasi} M.C. Gonz\'alez, C.A. Hidalgo, and A.-L. Barab\'asi, Nature {\bf 453}, 779--782 (2008).
\bibitem{condamin} S. Condamin, V. Tejedor, R. Voituriez, O. B\'enichou, and J. Klafter, Nature {\bf 450}, 77--80 (2007).
\bibitem{klafter} Y. Meroz, I.M. Sokolov, and J. Klafter, Phys. Rev. E {\bf 83}, 020104 (2011).
\bibitem{mantegna} J. P. Bouchaud and M. Potters, {\it Theory of Financial Risk} (Cambridge University Press, Cambridge, England, 2000).
\bibitem{lo} J.Y Campbell, A.W. Lo and A.C. MacKinlay, {\it The Econometrics of Financial Markets} (Pinceton University Press, 1997).
\bibitem{mp} J. Masoliver and J. Perell\'o, Phys. Rev. E {\bf 75}, 046110 (2007).
\bibitem{rutowski} T.R. Bielecki, M. Rutkowski, {\it Credit Risk} (Springer, New York, 2004)
\bibitem{alili} L. Alili and A. E. Kyprianou, Ann. Appl. Probab. {\bf 15}, 2062--2080 (2005)
\bibitem{avram} F. Avram, A.E. Kyprianou, and M.R. Pistorius, Ann. Appl. Probab. {\bf 14} 215--238 (2004).
\bibitem{chriss} F. Almgren, Applied Mathematical Finance {\bf 10}, 1--18 (2003).
\bibitem{hendershot} T. Hendershott, Ch.M. Jones, and A.J Menkveld, J. Finance {\bf 66}, 1--33 (2011).
\bibitem{masoliverfpt} J. Masoliver, J. Perell\'o, Phys. Rev. E {\bf 80}, 016108 (2009).
\bibitem{sazuka} N. Sazuka, J. Inoue, E. Scalas, Physica A {\bf 388}, 2839--2853 (2009).
\bibitem{havlin1} K. Yamasaki, L. Muchnik, S. Havlin, A. Bunde, and H. E. Stanley, Proc. Natl. Acad. Sci. U.S.A. {\bf 102}, 9424 (2005)
\bibitem{wang} F. Wang, K. Yamasaki, S. Havlin, and H. E. Stanley, Phys. Rev. E {\bf 79}, 016103 (2009).
\bibitem{lillo} M. Montero, J. Perell\'o, J. Masoliver, F. Lillo, S. Miccich\'e, R.N. Mantegna, Phys. Rev. E {\bf 72}, 056101 (2005).
\bibitem{bonano} G. Bonanno, D. Valenti, and B. Spagnolo, Eur. Phys. J. B {\bf 53}, 405--409 (2006)
\bibitem{valenti} D. Valenti, B. Spagnolo, and G. Bonanno, Physica A {\bf 382}, 311–-320 (2007).
\bibitem{bunde1} M.I. Bogachev, J.F. Eichner, and A. Bunde, Phys. Rev. Lett. {\bf 99}, 240601 (2007)
\bibitem{bogachev} M.I. Bogachev and A. Bunde, Phys. Rev. E {\bf 80}, 026131 (2009).
\bibitem{kantz} M. S. Santhanam and H. Kantz, Phys. Rev. E {\bf 78}, 051113 (2008).
\bibitem{ren} F. Ren and W.-X. Zhou, New Journal of Physics {\bf 12}, 075030 (2010).
\bibitem{klein} J.P. Klein and M. Moeschberger, {\it Survival analysis: techniques for censored and truncated data} (Springer, New York, 2003).
\bibitem{plerou} V. Plerou and H.E. Stanley, Phys. Rev. E {\bf 77}, 037101 (2008).
\bibitem{mantegna1} R.M. Mantegna and H.E. Stanley, Nature {\bf 376}, 46--49 (1995).
\bibitem{mmporra} J. Masoliver, M. Montero and J.M. Porr\`a, Physica A {\bf 283}, 559--567 (2000).
\bibitem{sparre} E. Sparre Andersen, Math. Scand. {\bf 1}, 263--285 (1953)
\bibitem{sparre1} E. Sparre Andersen, Math. Scand. {\bf 2}, 195--223 (1954).
\bibitem{chechkin} A.V. Chechkin, R. Metzler, V.Y. Gonchar, J. Klafter and L.V. Tanatarov, J. Phys. A: Math. Gen. {\bf 36}, L537--L544 (2003). 

\end{thebibliography}
\end{document}